\documentstyle[erice99,epsf,12pt]{article}

\newcommand{\be}[1]{\begin{equation} \label{(#1)}}
\newcommand{\ee}{\end{equation}}
\newcommand{\ba}[1]{\begin{eqnarray} \label{(#1)}}
\newcommand{\ea}{\end{eqnarray}}

\begin{document}
\title{Fixed-$t$ subtracted dispersion relations for Compton
  scattering off the nucleon 
\footnote{supported by Deutsche Forschungsgemeinschaft (SFB 443)}
}
\author{M. Gorchtein, D. Drechsel, B. Pasquini, M. Vanderhaeghen}
\address{Institut f\"ur Kernphysik, Johannes Gutenberg Universit\"at,
J.-J.-Becher-Weg 45, D-55099 Mainz, Germany}
%
\maketitle

\abstracts
{
Fixed-$t$ subtracted dispersion relations are presented for Compton
scattering off the nucleon at energies $E_\gamma\,\leq\,500$ MeV, as a
formalism to extract the nucleon polarizabilities with a minimum of
model dependence. The scalar polarizability difference $\alpha\,-\,\beta$
and the backward spin polarizability $\gamma_\pi$ enter directly as fit
parameters in this formalism.
}

{
\hspace{-1 truecm}
Compton scattering off the nucleon is determined by 6 independent
helicity amplitudes A$_i\,(i\,=\,1, \ldots ,6)$, which are functions
of two Lorentz invariants, $t$ and
$\nu\,=\,E_\gamma^{lab}\,+\,t/4$M. In the limit $\nu\,\rightarrow\,0$,
the leading term in the expansion in $\nu$ is determined by ground
state properties of the nucleon, i.e. its charge, mass and anomalous
magnetic moment. The internal structure of the nucleon shows up only
at order $\nu^2$ and can be parametrized in terms of polarizabilities.
\newline
Following the conventions of Ref. \cite{lvov}, we start with the
unsubtracted dispersion relations (DR) for the invariant amplitudes
A$_i(\nu,t)$ at fixed value of the variable $t$:

\vspace{-0.5 truecm}

\begin{equation}
\mathrm{ReA}_i(\nu,t)\,=\,\mathrm{A}_i^B(\nu,t)\,+\,\frac{2}{\pi}{\mathcal{P}}
\int_{\nu_{thr}}^\infty d\nu'\frac{\nu'
  \mathrm{ImA}_i(\nu',t)}{\nu'^2-\nu^2}\,,
\end{equation}

\hspace{-1.2 truecm} where A$_i^{\mathrm{B}}$ are the Born contributions and
$\nu_{thr}$is the threshold of pion 
photoproduction in the $s$-channel. Unfortunately for the two
amplitudes A$_1$ and A$_2$, the unsubtracted DR do not converge. In order
to avoid these convergency problems, we consider DR at fixed $t$ that
are once subtracted at $\nu\,=\,0$ \cite{my},

\vspace{-0.5 truecm}

\begin{equation}
\mathrm{ReA}_i(\nu,t)
\,=\,
\mathrm{A}_i^B(\nu,t)
\,+\,
\left[\mathrm{A}_i(0,t)\,-\,\mathrm{A}_i^B(0,t)\right]
\,+\,
\frac{2}{\pi} \nu^2 {\mathcal{P}}
\int_{\nu_{thr}}^\infty d\nu'
\frac
{\mathrm{ImA}_i(\nu',t)}
{\nu'(\nu'^2-\nu^2)}\,.
\end{equation}

\hspace{-1.2 truecm} The subtraction gives two additional powers of
energy in the denominator in (2), which ensures the convergence of all
6 dispersion integrals. The price to pay is the appearance of six
unknown subtraction functions, A$_i(0,t)\,-\,$A$_i^{\mathrm{B}}(0,t)$,
which we again determine by once-subtracted DR, this time in $t$
at fixed $\nu\,=\,0$:

\vspace{-0.8 truecm}

\begin{equation}
\mathrm{A}_i^{NB}(0,t)
\,=\,
\mathrm{A}_i^{NB}(0,0)
\,+\,\left[\mathrm{A}_i^{t-pole}(0,t)\,-\,\mathrm{A}_i^{t-pole}(0,0)\right]
\,+\,
\frac{t}{\pi}
\int_{t_{thr}}^\infty dt'
\frac
{\mathrm{ImA}_i(0,t')}
{t'(t'-t)}\,.
\end{equation}

\hspace{-1.2 truecm} Here
A$_i^{\mathrm{NB}}(\nu,t)\,\equiv\;$A$_i(\nu,t)\,-\,$A$_i^{\mathrm{B}}(\nu,t)$,
and $A_i^{\mathrm{t-pole}}(0,t)$ represents the contribution of poles in the
$t$-channel, in particular the $\pi^0$ pole in the case of A$_2$. 

\hspace{-1.2 truecm} The unitarity relation ImT = $\sum_X
T(i,X)\cdot T^{\ast}(f,X)$, where $i$ and $f$ are initial and final
states, and $X$ denotes all the possible intermediate states, relates
ImA$_i$ to the pion photoproduction multipoles 
(HDT analysis at E$_\gamma^{lab}\leq$ 500MeV and SAID solution SP98K
from 500MeV to 1.5GeV) in the $s$-channel 
and to the amplitudes for the subprocesses 
$\gamma\gamma\,\rightarrow\,\pi\pi$ and $\pi\pi\,\rightarrow\,N
\bar{N}$ in the $t$-channel. For $\gamma\gamma\,\rightarrow\,\pi\pi$
we take the Born amplitude with unitarized $s$ and $d$-waves and for 
$\pi\pi\,\rightarrow\,N \bar{N}$ the extrapolation of the $\pi$N
scattering amplitude in the unphysical region \cite{hoehler}.
Thanks to the subtraction, the dispersion integrals converge quite fast.
Thus, the higher energy contributions are small and can be estimated
by simple models (e.g. the resonant part of $\pi\pi$N intermediate
states is included in the $s$-channel integral).
The subtraction constants $A_i^{\mathrm{NB}}(0,0)$ are directly related to the
nucleon polarizabilities. Four of them (i = 3,$\ldots$,6) can be
calculated from the unsubtracted dispersion relations (1) at
$\nu=t=0$. The other two, $A_1^{\mathrm{NB}}(0,0)\sim (\alpha - \beta)$
and $A_2^{\mathrm{NB}}(0,0)\sim \gamma_\pi$, are used as fit
parameters in the present formalism. The
forward scalar polarizability $\alpha + \beta$ is fixed to 13.69$\cdot
10^{-4}fm^3$ \cite{babusci}. 
In Fig.1 we show the c.m. differential cross section as function of
the c.m. scattering angle at the lab photon energy 182 MeV in
comparison with the experimental data of Ref. \cite{data}.

\vspace{-5 truecm}
\begin{figure}[ht]
\epsfxsize = 14cm
\epsfysize = 18cm
\centerline{\epsffile {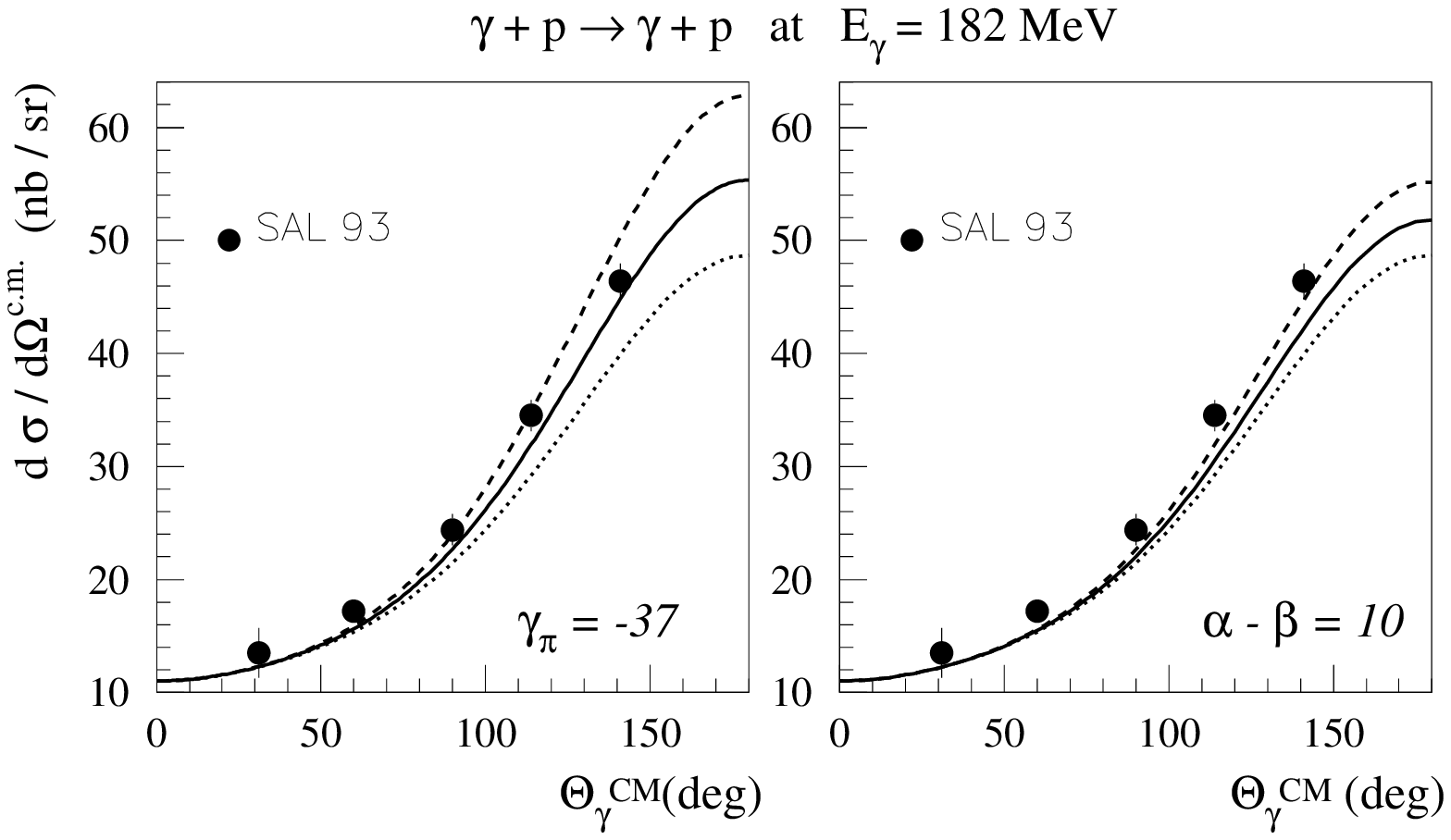} }
\vspace{-6.5 truecm}
\caption[] 

The left picture displays the sensitivity of the angular distribution
to varying $\alpha - \beta$ around 8 (in units of
$10^{-4}fm^3$) at fixed $\gamma_\pi\,=\,-37 $(in units of
$10^{-4}fm^4$), results are shown for $\alpha - \beta\,=\,8$ (solid line),
10 (dashed line), and 6 (dotted line).
The right picture displays the sensitivity of the angular distribution
to varying $\gamma_\pi$ at fixed $\alpha - \beta\,=\,10$, results are
shown for $\gamma_\pi\,=\,-32$ (solid line), -27 \cite{sandorfi}
(dashed line), and -37 \cite{gamma} (dotted line).

\end{figure}
\vspace{-1.3 truecm}


}


\end{document}